\documentclass[11pt]{article}
\usepackage{graphicx}
\usepackage{latexsym}
\usepackage{amsfonts}

\newcommand{\qed}{\hfill $\Box$}

\newtheorem{theorem}{Theorem}[section]
\newtheorem{lemma}[theorem]{Lemma}
\newtheorem{proposition}[theorem]{Proposition}
\newtheorem{corollary}[theorem]{Corollary}

\newtheorem{example}[theorem]{Example}
\newtheorem{assumption}[theorem]{Assumption}

\newtheorem{remark}[theorem]{Remark}

\begin{document}

\title{Smooth Value Function with Applications in Wealth-CVaR Efficient Portfolio and Turnpike Property}
\author{Baojun Bian\thanks{Department of Mathematics,
         Tongji University, Shanghai 200092, China.
bianbj@tongji.edu.cn, Research of this author was supported by
NSFC 11071189 and 71090404.}\   and
Harry Zheng\thanks{Department of Mathematics, Imperial College, London SW7 2BZ, UK.
h.zheng@imperial.ac.uk.}}

\date{}
\maketitle

\noindent{\bf Abstract.}
In this paper we continue the study of Bian-Miao-Zheng (2011) and extend the results there to a more general class  of utility functions which may be bounded and non-strictly-concave and show that there is a classical solution to the HJB equation with the dual control method. We then apply the results to study  the efficient frontier  of wealth and conditional VaR (CVaR) problem and the turnpike property problem. For the former we construct explicitly  the optimal control and discuss the choice of the optimal threadshold level  and illustrate  that the wealth and the CVaR are positively correlated. For the latter we give a simple proof to the turnpike property of the optimal policy of long-run investors  and generalize the results   of Huang-Zariphopoulou (1999).

\medskip\noindent{\bf Key words.}
Non-strictly-concave utility function, HJB equation, dual control,
smooth value function, efficient frontier of wealth and CVaR, turnpike property.

\medskip\noindent{\bf AMS subject classifications.} 90C46, 49L20

%==================================================
% section 1
%==================================================

\section{Introduction}
One of the standard methods for utility maximization  is stochastic control which applies the dynamic programming principle and Ito's lemma to derive a nonlinear parabolic partial differential equation (PDE), called the HJB equation, for the optimal value function. If there is a   smooth solution  one may then apply the verification theorem to show that the value function is a classical solution to the HJB equation and find the optimal control as a byproduct.  For excellent expositions of stochastic control and its applications in utility maximization, see \cite{fs,pham} and references therein.

The smoothness of the value function is a highly desirable property. One normally has to impose some conditions to ensure that. One key condition is the uniform ellipticity of the diffusion coefficient, which is not
 satisfied for the standard wealth process as long as doing nothing is a feasible portfolio trading strategy.
When the trading constraint set is a closed convex cone,
the utility function  is strictly concave, continuously differentiable, satisfying some growth conditions,  and the market is complete, the value function is a classical solution to the HJB equation, see \cite{ks98}. When the constraint set is the whole space and
the utility function is of  power, logarithmic, exponential type, the value function has a  closed-form expression.

For general continuous increasing concave functions $U$ (not necessarily continuously differentiable or strictly concave) satisfying $U(0)=0$ and $U(\infty)=\infty$  \cite{bmz11}  shows that there exists a smooth solution to the HJB equation and that the value function is a classical solution if some exponential moment condition is satisfied.
The key idea is to find a smooth solution to the dual HJB equation and then to show the conjugate function of the dual solution is a smooth solution to the primal HJB equation.

In this paper we first extend the regularity of the value function, similar to that in \cite{bmz11}, to more general utility functions by removing the requirement $U(\infty)=\infty$. We show that the value function is continuous and is a constant if state variable $x$ is above a fixed threshold level and a smooth solution to the HJB equation if $x$ is below the threshold level with a representation in terms of the dual value function satisfying the dual HJB equation. The smoothness property of the value function not only provides a feedback optimal control and a classical solution to the HJB equation but also enables us to study other  related problems. We discuss in detail two applications: One is the efficient frontier of wealth  and CVaR, the other is the turnpike property of  the optimal policy of a long run investor.

The utility function of  terminal wealth maximization  up to a positive constant level $H$ is given by
$U(x)=x\wedge H =\min(x,H)$ for $x\geq 0$ and $-\infty$ for $x<0$, which is not covered by \cite{bmz11} as $U(\infty)=H$ and not $\infty$.
The problem  is studied in \cite{x04}  in a complete market model where the martingale method is used to first solve a static optimization problem and find the optimal terminal wealth, and then to compute the optimal value function  and verify it satisfies the HJB equation. We derive the same result of \cite{x04} in this paper by applying Theorem \ref{maintheorem} to a specific utility function $U(x)=x\wedge H$ and  Theorem \ref{verification} for verification of optimality. We go further to discuss the associated CVaR, the optimal choice of the level $H$, and the efficient frontier of wealth and CVaR with a two-stage optimization approach.

The turnpike property  is a classic problem in finance and has been discussed by many researchers for both discrete time and continuous time models. Cox and Huang \cite{ch92} use the probabilistic method to show that the portfolio turnpike property holds if the inverse of the marginal utility function satisfies some conditions. Huang and Zariphopoulou \cite{HZ99} establish the turnpike property with the viscosity solution method to the HJB equation when the marginal utility function behaves like that of a power utility at a large wealth level. Using the dual relation of the primal and  dual value functions, we derive the same turnpike property and give a direct and simple proof. The conditions on the utility function are also relaxed. One sufficient condition for the turnpike property is that the utility function behaves like a power utility at a large wealth level and some technical conditions on second order differentiability of the utility function and the convexity of the marginal utility function are not required.

We have made several contributions in this paper: We extend the results of \cite{bmz11} to more general utility functions, construct a closed form solution to the HJB equation for a wealth maximization problem, demonstrate the positive relation between the wealth and the risk, and derive the turnpike property for a broad class of utility functions.

The rest of the paper is organized as follows.  Section 2 describes the model formulation and constructs a
classical solution to the  HJB equation via a smooth dual value function  and shows the primal value function is smooth with the verification theorem
under an exponential moment condition for the optimal control. Section 3 studies a  portfolio wealth maximization problem, derives the closed form optimal solution, and discusses the related issues such as CVaR, the choice of threshold level, and the efficient frontier. Section 4 focuses on the turnpike property and shows that for a broad class of utility functions the optimal policy for long run investors is to invest a constant proportion of the wealth in  risky assets. Section 5 concludes.

%====================================================
% section 2
%====================================================
\section{Smooth HJB Solutions and Value Functions}
Consider a financial market consisting of one bank account  and $n$ stocks.
The  price process  $S=(S^1,\ldots,S^n)^T$ of $n$ risky assets
is  modelled by
$$
dS_t={\rm diag}(S_t)(\mu(t) dt+ \sigma(t) dW_t),\quad 0\leq t\leq T
$$
with the initial price $S_0=s$,
where  $x^T$ is the transpose of $x$, ${\rm diag}(S_t)$ is an $n\times n$ matrix with diagonal elements $S^i_t$
and all other elements zero,
$\mu$ and $\sigma$ are deterministic continuous
vector-valued and nonsingular matrix-valued functions of time $t$,
representing stock returns and volatilities, respectively,  and
 $W$ is  an $n$-dimensional standard Brownian motion
 on a complete probability space
$(\Omega,\mathcal{F},P)$, endowed with a natural filtration
$\{\mathcal{F}_t\}$ generated by $W$, augmented by all $P$-null sets.
The riskless interest rate is a positive constant denoted by $r$.
The  wealth process $X$  satisfies the stochastic differential equation (SDE)
\begin{equation} \label{wealth}
dX_t = X_t[(\pi_t^T (\mu(t)-r{\mathbf 1}) +r)dt + \pi_t^T\sigma(t) dW_t],
\quad X_0=x,
\end{equation}
where ${\bf 1}$ is a vector with all components 1 and $\pi_t$ are progressively measurable
control processes satisfying    $\pi_t\in K$, a closed convex
    cone in $R^n$, a.s. for $t\in[0,T]$ a.e. $\pi_t$ represent proportions of wealth $X_t$ invested in risky assets $S_t$. In our notation we write time $t$ in parentheses for deterministic functions (e.g., $b(t),\sigma(t)$) and in subscript for stochastic processes (e.g.,  $S_t,\pi_t$).

A terminal wealth utility maximization problem is defined by
\begin{equation} \label{primal}
\sup_{\pi} E[U(X_T)]
\mbox{ subject to (\ref{wealth})},
\end{equation}
where $U$ is a utility function satisfying the following conditions.
\begin{assumption} \label{utility}
$U$ is a continuous increasing and  concave function on $[0,\infty)$,
satisfying $U(0)>-\infty$ and
\begin{equation}\label{growth}
 U(x)\leq  C(1+x^p),\ x\geq 0
\end{equation}
for some constant  $C>0$ and $0<p<1$.
\end{assumption}

Assumption \ref{utility} is satisfied by power utility $U(x)={1\over p}x^p$ with $0<p<1$,
  exponential  utility $U(x)=-e^{-\alpha x}$ with $\alpha>0$, and other utilities such as $U(x)=x\wedge H$ with $H>0$.  We denote by $C$ a generic positive constant.

Denote by $V(t,x)$ the  value function of
(\ref{primal}) for $0\leq t\leq T$ and $x\geq0$, defined by
\[
V(t,x)=\sup_{\pi} E[U(X_T)|X_t=x].
\]
The  HJB equation is given by
\begin{equation} \label{HJB}
 -{\partial V\over \partial t}(t,x) - \sup_{\pi\in K}
\{(\pi^Tb(t)+r)xV_x(t,x) + {1\over 2}|\sigma(t)^T\pi|^2x^2 V_{xx}(t,x)\}
 =0,\; x>0,t<T
\end{equation}
with the terminal condition $V(T,x)=U(x)$, where $b(t)=\mu(t)-r{\bf 1}$ is the stock excess return, 
${\partial V\over \partial t}$ the partial derivative of $V$ with respect to $t$, $V_x$ and $V_{xx}$  similarly defined. If we define $\bar V(t,x)=V(t,e^{r(T-t)}x)$, then $\bar V$ satisfies
\begin{equation} \label{HJBa}
 -{\partial\bar V\over \partial t}(t,x) - \sup_{\pi\in K}
\{\pi^Tb(t)x\bar V_x(t,x) + {1\over 2}|\sigma(t)^T\pi|^2x^2 \bar V_{xx}(t,x)\}
 =0,\; x>0,t<T.
\end{equation}
Since the regularity properties of $V$ and $\bar V$ are the same and the HJB equation (\ref{HJBa}) corresponds to the case $r=0$ (or the discounted stock price process), we will focus on (\ref{HJBa}) in the remaining part of the section and continue to write $V$ instead of $\bar V$.  All regularity results derived apply to the HJB equation (\ref{HJB}) for  $r>0$.

The dual function of $U$  is defined by
$$ \tilde U(y)=\sup_{x\geq 0} (U(x)-xy).$$
The function $\tilde U$ is a continuous decreasing and convex function on $[0,\infty)$, satisfying   $\tilde U(\infty)=U(0)$ and
\begin{equation} \label{dualgrowth}
\tilde U(y)
\leq C(1+y^{q}),\ y>0
\end{equation}
for some constant  $C>0$ and $q={p\over p-1}<0$.

\begin{remark}{\rm
Compared with \cite{bmz11} we have removed the  condition  $U(\infty)=\infty$, which is not satisfied for utility functions with upper bound.  Since all other conditions are the same as those in \cite{bmz11} most results in that paper still hold in the current setting. We  state the key results but only prove the parts which are different and refer the reader to  \cite{bmz11} for detailed proofs of all other parts.
}\end{remark}

The dual process $Y$ satisfies the SDE
\begin{equation} \label{dualSDE}
dY_t=-Y_t(\sigma(t)^{-1}\nu_t+\theta(t))^T dW_t, \quad Y_0=y,
\end{equation}
where $\nu$ are progressively measurable control processes satisfying $\nu_t\in\tilde{K}$, the positive polar cone of $K$ in $R^n$,  a.s. for  $t\in [0,T]$ a.e. and $\theta(t)=\sigma(t)^{-1}b(t)$.

The dual problem is defined by
$$ \inf_\nu E[\tilde U(Y_T)]\mbox{ subject to (\ref{dualSDE})}.
$$

Denote by $v(t,y)$ the dual value function, i.e.,
$v(t,y)=\inf_\nu E[\tilde U(Y_T)|Y_t=y]$.
The dual HJB equation is a linear PDE
\begin{equation} \label{dualHJB}
{\partial v\over \partial t} (t,y)
+\frac{1}{2}|\hat{\theta}(t)|^2 y^2v_{yy}(t,y)=0,\
y>0,\; 0\leq t<T
\end{equation}
with the terminal condition $v(T,y)= \tilde U(y)$,
where $\hat{\theta}(t)=\theta(t)+\sigma(t)^{-1}\hat{\pi}(t)$ and
$\hat{\pi}(t)$ is the unique
minimizer of $f(\tilde{\pi})=|\theta(t)+\sigma(t)^{-1}\tilde{\pi}|^2$ over $\tilde{\pi}\in \tilde{K}$.

\begin{assumption}
\label{parabolicity} $\hat{\theta}$
is continuous on $[0,T]$  and there is a positive
constant $\theta_0$ such that $|\hat{\theta}(t)|\geq \theta_0$
for all $t\in [0,T]$.
\end{assumption}

\begin{remark}\label{rk3.2a}
{\rm
Assumption \ref{parabolicity} is automatically satisfied if all components of $b(t)$ are positive, a natural condition as $b(t)$ represents
the stock excess returns, and $K$ is either the whole space $R^n$ (no trading constraints) or the nonnegative part of the whole space $R^n_+$ (short selling constraints). The positive polar cone $\tilde K$ is then either $\{0\}$ or $R^n_+$ and the optimal solution $\hat\pi(t)=0$ for all $t$. Therefore $\hat\theta(t)=\theta(t)$ is a nonzero continuous vector-valued function on $[0,T]$ and $\theta_0$ is the minimum value of
$|\theta|$ on $[0,T]$.
}\end{remark}

\begin{lemma} \label{LemhatV}
The  function $v$ is continuous on $[0,T]\times (0,\infty)$ and
 is a classical solution to (\ref{dualHJB}), satisfying
\[
U(0)\leq v(t,y)\leq C (1+y^{q}),\ t\in[0, T], y>0
\]
for some  constant  $C>0$. Furthermore, for every $t\in[0, T)$,
 $v(t,\cdot)$ is strictly decreasing and strictly convex with the following limiting properties:
\begin{equation} \label{limits}
\lim_{y\rightarrow 0} v(t,y)=\tilde U(0),\quad \lim_{y\rightarrow
\infty}v(t,y)=U(0),\quad
\lim_{y\rightarrow 0}v_y(t,y)=\tilde U'(0), \quad \lim_{y\rightarrow
\infty}v_{y}(t,y)=0,
\end{equation}
where $\tilde U'$ is the right directional derivative of $\tilde U$.
\end{lemma}

\noindent {\it Proof.} We only need to prove (\ref{limits}) as all other parts have been proved in \cite[Lemmas 3.5 and 3.6]{bmz11}.
Note that
$$ v(t,y)=(2\sqrt{\pi\tau})^{-1} \int_{-\infty}^\infty e^{-{1\over 4\tau}x^2}
\tilde U(ye^{x-\tau}) dx,$$
where $\tau=\frac{1}{2}\int_t^T |\hat{\theta}(\eta)|^2 d\eta $. Since
 $\tilde U(ye^{x-\tau})$ is bounded below by $U(0)$ and  is increasing as $y\to 0$, the Monotone Convergence Theorem (MCT) confirms the first limit in (\ref{limits}).

To show the second limit in (\ref{limits}), we may say $y>1$ and get $y^{q}<1$ as $q<0$, which gives  $U(0)\leq \tilde U(ye^{x-\tau})\leq C(1+e^{q(x-\tau)})$,  the Dominated Convergence Theorem (DCT)  implies the required limit.

To show the third limit in (\ref{limits}), we write
$$ {v(t,y+h)-v(t,y)\over h}
=(2\sqrt{\pi\tau})^{-1} \int_{-\infty}^\infty e^{-{1\over 4\tau}(x-2\tau)^2}
g(x,y; h) dx,$$
where
$$ g(x,y;h)={\tilde U(ye^{x-\tau}+he^{x-\tau}) - \tilde U(ye^{x-\tau})
\over he^{x-\tau}}
$$
which converges to $\tilde U'(ye^{x-\tau})$ 
as $h\downarrow 0$. Since $V$ is convex and decreasing and $g$ is
increasing with respect to $h$, we have $g(x,y;h)$ is decreasing as $h\downarrow
0$ and $g(x,y;h)\leq 0$ for all $h\geq 0$. The MCT says that
\begin{eqnarray*}
v_y(t,y) &=& \lim_{h\downarrow 0}  {v(t,y+h)-v(t,y)\over h}\\
&=& (2\sqrt{\pi\tau})^{-1} \int_{-\infty}^\infty e^{-{1\over 4\tau}(x-2\tau)^2}
\lim_{h\downarrow 0}g(x,y; h) dx\\
&=& (2\sqrt{\pi\tau})^{-1} \int_{-\infty}^\infty e^{-{1\over 4\tau}(x-2\tau)^2}
\tilde U'(ye^{x-\tau}) dx.
\end{eqnarray*}
Since $V$ is convex, we have $\tilde U'(y)$ is increasing with respect
to $y$ and $\tilde U'(ye^{x-\tau})\leq \tilde U'(\infty)=0$ and $\tilde U'(ye^{x-\tau})$ is decreasing as $y\to 0$. Applying the MCT again we get
$$ v_y(t,0)=\lim_{y\to 0}v_y(t,y)
= (2\sqrt{\pi\tau})^{-1} \int_{-\infty}^\infty e^{-{1\over 4\tau}(x-2\tau)^2}
\lim_{y\to 0} \tilde U'(ye^{x-\tau}) dx
= \tilde U'(0).$$

Finally, to show the fourth limit in (\ref{limits}), we note that
$$ {v(t,y)\over y}= (2\sqrt{\pi\tau})^{-1} \int_{-\infty}^\infty e^{-{1\over 4\tau}(x-2\tau)^2} {\tilde U(ye^{x-\tau})\over ye^{x-\tau}} dx.$$
For $y\geq 1$, we have
$$ {\tilde U(ye^{x-\tau})\over ye^{x-\tau}} \geq {U(0)\over ye^{x-\tau}} $$
which is bounded below by 0 if $U(0)$ is nonnegative or by $U(0)e^{-x+\tau}$ if $U(0)$ is negative.  On the other hand,
$${\tilde U(ye^{x-\tau})\over ye^{x-\tau}}
\leq C{(1+y^{q}e^{(x-\tau)q})\over y e^{x-\tau}}
\leq C{(1+e^{(x-\tau)q})\over e^{x-\tau}}.$$
Applying the DTC we get
$$ v_y(t,\infty)=\lim_{y\to\infty}{v(t,y)\over y}=
(2\sqrt{\pi\tau})^{-1} \int_{-\infty}^\infty e^{-{1\over 4\tau}(x-2\tau)^2} \lim_{y\to\infty} {\tilde U(ye^{x-\tau})\over ye^{x-\tau}} dx
=0.
$$
We have proved all limits in (\ref{limits}).
 \qed

We  now construct a classical solution to the HJB equation (\ref{HJBa}).
\begin{theorem}
\label{maintheorem}
 Assume $K$ is a closed convex cone and
Assumptions \ref{utility} and \ref{parabolicity} hold. Then there
exists a  function  $u\in C^0([0,T]\times
[0,\infty))$ which is a
classical solution to the HJB equation (\ref{HJBa}) in the region
$[0,T)\times (0,-\tilde U'(0))$ and has the following representation
$$
u(t,x)=\left\{\begin{array}{ll}
v(t, y(t,x)) + xy(t,x), & 0\leq x< -\tilde U'(0) \\
\tilde U(0), & x\geq -\tilde U'(0),
\end{array}\right.
$$
where $y\in
C^{1,\infty}([0, T)\times (0,-\tilde U'(0)))$ satisfies
$$ v_y(t, y(t,x))+x=0.$$
Furthermore, for $0\leq x< -\tilde U'(0)$ the maximum of
the Hamiltonian in the HJB equation (\ref{HJBa}) is achieved at
\begin{equation} \label{control}
\pi^*(t,x)=-(\sigma(t)^T)^{-1}\hat\theta(t){u_x(t,x)\over x u_{xx}(t,x)}
\end{equation}
 and $\pi^*(t,x)\in K$, the function
$u(t,x)$ is strictly increasing and strictly concave in $x$ for
fixed $t\in [0,T)$ and satisfies $u(T,x)=U(x)$ and $
u(t,x)\leq C(1+x^p)$ for some  constant $C>0$.
\end{theorem}

\noindent {\it Proof.} For $(t,x)\in [0,T]\times R_+$ define
\begin{equation} \label{u}
u(t,x)= \inf_{y> 0}\{v(t,y)+xy\}.
\end{equation}
If $x\geq -\tilde U'(0)$ then
$$ v(t,y)\geq v(t,0) +v_y(t,0)y = \tilde U(0) + \tilde U'(0) y \geq \tilde U(0) -xy$$
Therefore
$$ v(t,y)+xy\geq \tilde U(0)=v(t,0)+x0.$$
The minimum point is $y^*=0$.

If $0<x<-\tilde U'(0)$ then minimum is achieved at a point $y$ satisfying
$v_y(t,y)+x=0$.
Let $y(t,\cdot)$ be the inverse function of $-v_y(t,\cdot)$,
i.e.,
$$
-v_y(t,y(t,x))= x,\ y(t,-v_y(t,y))=y,
$$
for fixed $t\in[0, T)$. $y(t,x)$ is well defined on
$[0,T)\times(0,-\tilde U'(0)) $ from Lemma \ref{LemhatV}.
Since $v\in C^{1,\infty}([0, T)\times
(0,\infty))$ and $v_{yy}(t,y)>0$, the inverse function $y\in
C^{1,\infty}([0, T)\times (0,-\tilde U'(0)))$ by  the Implicit
Function Theorem. We have, for $(t,x)\in[0,T)\times(0,\infty)$,
\begin{eqnarray*}
u(t,x)&=&\left\{
\begin{array}{ll}
v(t, y(t,x)) + xy(t,x), & 0<x<-\tilde U'(0)\\
v(t,0), & x\geq  -\tilde U'(0).
\end{array}\right.
\end{eqnarray*}
which yields the regularity of $u(t,x)$. Note that if $x=-\tilde U'(0)$ then
$-v_y(t, y(t,x))=-\tilde U'(0)$ which implies $y(t,x)=0$. So $u(t,\cdot)$ is continuous at $x=-\tilde U'(0)$ (if $\tilde U'(0)$ is finite).

For $0<x<-\tilde U'(0)$, since $y(t,x)>0$ and $v_{yy}(t,y(t,x))>0$ for fixed $0\leq
t<T$, the function $u(t,\cdot)$ is strictly increasing and
strictly concave.  A direct computation yields
$$
{\partial u\over \partial t}(t,x)
-\frac{1}{2}|\hat{\theta}(t)|^2\frac{u_x^2(t,x)}{u_{xx}(t,x)}=0.
$$
We conclude by  \cite[Lemma 3.7]{bmz11}
 that $u$ is a classical solution to the
HJB equation (\ref{HJBa}) and the maximum of the Hamiltonian is achieved at $\pi^*(t,x)$.  Furthermore, from Lemma \ref{LemhatV} we have
$u(t,x)\leq  C(1+x^p)$
for some constant $C>0$.
\qed

\begin{remark}%\label{rk3.2}
{\rm In \cite{bmz11} $U$ satisfies $U(\infty)=\infty$ which implies
$\tilde U(0)=\infty$ and $\tilde U'(0)=-\infty$.  The case $x\geq -\tilde U'(0)$ cannot happen and $u(t,x)$ is a classical solution to the HJB equation. If $\tilde U(0)<\infty$ we may have $\tilde U'(0)=-\infty$ (e.g., $U(x)=-e^{-\alpha x}$, $\tilde U(0)=0$ and $\tilde U'(0)=-\infty$) or $\tilde U'(0)>-\infty$ (e.g., $U(x)=x\wedge H$, $\tilde U(0)=H$ and
$\tilde U'(0)=-H$).
}\end{remark}

\begin{remark}
{\rm  In  fact,  function $u\in C^0([0,T]\times
[0,\infty))$ is a
viscosity solution to the HJB equation (\ref{HJBa}) in the region
$[0,T)\times (0,\infty)$. We only need to verify that $u$ satisfies the viscosity inequalities
on $(-\tilde U'(0),t_0)$ for $t_0\in [0,T)$ if $\tilde U'(0)>-\infty$. Denote by $x_0=-\tilde U'(0)$. It is easy to see that
the second-order superjet of $u$ at point $(x_0,t_0)$ is given by
\[
{J^{2,+}}u(t_0,x_0)=\{(q,p,A)|q\geq 0,p=0,A\geq 0\}
\]
if $u$ is differentiable at $(t_0,x_0)$ and 
\begin{eqnarray*}
{J^{2,+}}u(t_0,x_0)&=&\{(q,p,A) |q\geq 0,p=\{0,u^-_x(t_0, x_0)\},A\geq 0\}\\
&& \bigcup \{(q,p,A)|q\geq 0,0<p<u^-_x(t_0, x_0), -\infty<A<\infty\}
\end{eqnarray*}
if $u$ is not differentiable at $(t_0,x_0)$, where  $u^-_x(t_0, x_0)$ is the left directional derivative of $u$ with respect to $x$ at point $(t_0, x_0)$ and in this case $u^-_x(t_0, x_0)>0$.
The second-order subjet of $u$ at point $(t_0,x_0)$ is given by
\[
{J^{2,-}}u(t_0,x_0)=\{(q,p,A)|q\leq 0,p=0,A\leq 0\}
\]
if $u$ is differentiable at $(t_0,x_0)$ and
\[
{J^{2,-}}u(t_0,x_0)=\emptyset
\]
if $u$ is not differentiable at $(t_0,x_0)$.
These imply that $u$ is a viscosity solution.
}\end{remark}

Theorem \ref{maintheorem} confirms that there is a classical
solution $u$ to the HJB equation (\ref{HJBa}) for $0<x<-\tilde U'(0)$ and the Hamiltonian achieves its
maximum at a point $\pi^*$ in $K$. The verification theorem next shows  that the value function $V$ is indeed a smooth classical solution
to the HJB equation (\ref{HJBa}) with the optimal feedback control $\pi^*$.

\begin{theorem} \label{verification}
Let $u$ be given as in Theorem \ref{maintheorem},  then the value function $V(t,x)\leq u(t,x)$ on $[0,T]\times (0,\infty)$. Furthermore, if  $x\geq -\tilde U'(0)$ then $V(t,x)=u(t,x)=\tilde U(0)$ and the optimal control is $\pi^*\equiv 0$; if $0<x<-\tilde U'(0)$, and
SDE (\ref{wealth})
admits a unique nonnegative strong solution $\bar X$
with the feedback control $\pi^*$ defined in (\ref{control})
and $\pi^*$ satisfies an exponential moment condition
\begin{equation} \label{novikov}
 E\left[\exp\left({1\over 2}\int_0^T |\pi^*(s,\bar X^{t,x}_s)^T\sigma(s)|^2 ds\right)\right]< \infty,
\end{equation}
then $V(t,x)=u(t,x)$ on $[0,T]\times [0,\infty)$ and $\pi^*$ is an optimal Markovian control.
\end{theorem}

\noindent{\it Proof}.\
For $x\geq -\tilde U'(0)$ we have $u(t,x)=\tilde U(0)=U(\infty)$. If we take a trivial control $\pi^*(t,x)\equiv 0$,  then $E[U(X_T)]=U(x)$.
Note that, for any $y>0$,
$$ \tilde U(y)\geq \tilde U(0) + \tilde U'(0)y \geq \tilde U(0) -xy$$
therefore
$$ \tilde U(0)\leq \inf_{y>0}(\tilde U(y) +xy) = U(x)$$
but $\tilde U(0)=U(\infty)$ so
$$ E[U(X_T)]\geq U(x)\geq U(\infty)\geq V(t,x)$$
which implies $X_T$ is the optimal terminal wealth when $\pi^*\equiv 0$ and $V(t,x)=U(x)=\tilde U(0)$.

For $0<x< -\tilde U'(0)$ we can use Ito's lemma  to show that
$V(t,x)\leq u(t,x)$ and apply the localization method and uniform integrability to show that $V(t,x)=u(t,x)$, see the detailed proof in \cite{bmz11}, Theorem 4.1.
\qed

\begin{remark}{\rm
We can in fact show, with the same method, that there exists a classical solution $u$ to the HJB equation if the growth condition (\ref{growth}) for utility $U$ is replaced by
\begin{equation} L_1x^q\leq U(x)\leq L_2(1+x^p),\quad x>0 \label{lowerbound}
\end{equation}
for some constants $L_1<0$, $L_2>0$, $q<0$ and $0<p<1$. Condition (\ref{lowerbound}) covers logarithemic utility $U(x)=\ln x$ and negative power utility $U(x)={1\over \gamma}x^\gamma$ with $\gamma<0$.  The solution $u$ satisfies
\begin{equation} \tilde L_1 x^q\leq u(t,x)\leq \tilde L_2 (1+x^p) \label{lowerbound_u}
\end{equation}
for some constants $\tilde L_1<0$ and $\tilde L_2>0$. However, it is not clear the growth condition (\ref{lowerbound_u}) is strong enough to get the uniform integrability property of a localization sequence, which is needed to prove the verification theorem with Ito's lemma (see details in \cite{bmz11}). We  cannot claim that the value function is a smooth solution to the HJB equation for $U$ satisfying (\ref{lowerbound}). This is still an open question and requires further research.
}\end{remark}

\begin{remark}{\rm
The function $v(t,y)$ is $C^{1,\infty}$ by its construction. The Implicit Function
Theorem then implies that $Y(t,x)$ and $u(t,x)$ are
also $C^{1,\infty}$ for $0<x<-\tilde U'(0)$. Using the same proof as \cite[Theorem 6.4.7]{pham} we can show that
if $\pi$ and $X$ are the optimal control and  state
processes and if we define
$$ (Y_t,Z_t)=(u_x(t, X_t), u_{xx}(t, X_t) X_t \pi_t^T \sigma(t) )$$
then $(Y_t,Z_t)$ are the adapted solution to the adjoint linear backward SDE
$$ -dY_t= (b(t)^T\pi_t Y_t + {\rm Tr}(\sigma(t)^T\pi_t Z_t)) dt - Z_t dW_t,$$
where ${\rm Tr}(A)$ is the trace of a square matrix $A$.
Since $U$ is not differentiable, the terminal condition should be modified
as
$$ Y_T\in \partial U(X_T),$$
where $\partial U(x)$ is the superdifferential of $U$ at $x$. The following
maximum condition is satisfied
$$ G(t,X_t,\pi_t, Y_t, Z_t)= \max_{\pi\in K}
G(t,X_t,\pi, Y_t, Z_t),$$
where
$$ G(t,x,\pi,Y,Z)=xb(t)^T\pi Y + {1\over 2}x{\rm Tr}(
\sigma(t)^T\pi Z).
$$
This is the stochastic maximum principle from the dynamic programming
and the HJB equation. Note that $X_t$ does not appear in the  backward SDE for $(Y_t,
Z_t)$ due to the linear structure of the SDE of the wealth process
$X_t$.
}\end{remark}

%==================================================
% wealth risk efficient frontier
%==================================================
\section{Efficient Frontier of Wealth and CVaR}
In this section we discuss the wealth maximization problem (\ref{primal}) and the efficient frontier of wealth and CVaR  problem, see \cite{zheng09} for  details on the existence of optimal solutions. Since both  involve  piecewise linear utility functions, we first study general piecewise linear utility functions and then apply the results to specific cases.

Assume $U$ is a piecewise linear function given by
\begin{equation}
 U(x)=\sum_{i=0}^N (c_{i+1} x + d_{i+1})1_{\{x_{i}\leq x< x_{i+1}\}}, \label{piecewise}
\end{equation}
where $N$ is a positive integer, $0=x_0<x_1<\cdots<x_N<x_{N+1}=\infty$, and $\infty=c_0>c_1>\cdots>c_N>c_{N+1}=0$. The continuity of  $U$ implies that
$$ c_ix_i+d_i=c_{i+1}x_i+d_{i+1}$$
for $i=1,\ldots,N$, that is, given $c_1,\dots,c_N$ and $d_{N+1}$ other coefficients $d_i$, $i=1,\ldots,N$, are uniquely determined. The utility function $U$ is an increasing  concave piecewise linear function which is non-differentiable at points $x_i$, $i=1,\ldots,N$,  non-strictly concave,
$U(0)=d_{1}, U(\infty)=d_{N+1}$ and $U'(0)=c_1, U'(\infty)=0$.

\begin{lemma} Let $U$ be defined in (\ref{piecewise}). Then the dual function of
$U$ is given by
\begin{equation}
\tilde U(y)=\sum_{i=0}^N (-x_i y+c_{i+1}x_i +d_{i+1}) 1_{\{c_{i+1}\leq y<c_i\}}. \label{dual-piecewise}
\end{equation}
The dual function $V$ is a decreasing convex piecewise linear function which is non-differentiable at points $c_i$, $i=1,\ldots,N$,  non-strictly convex,  $\tilde U(0)=d_{N+1}, \tilde U(\infty)=d_1$ and $\tilde U'(0)=-x_N, \tilde U'(\infty)=0$.
\end{lemma}

\noindent{\it Proof}.
Note that
\begin{eqnarray*} \tilde U(y)
&=& \sup_{x\geq 0} (U(x)-xy)\\
&=&\max_{i=1,\ldots,N+1} \max_{x_{i-1}\leq x< x_i} ((c_i - y)x + d_i)\\
&=& \max_{i=1,\ldots,N+1}  ((c_i-y)(x_{i-1}1_{\{c_i-y<0\}}
+ x_i 1_{\{c_i-y\geq 0\}}) +d_i).
\end{eqnarray*}
Clearly we need to know the position of $y$ for further discussion. Since $0=c_{N+1}<c_N<\cdots< c_1<c_0=\infty$ we have that for $y\geq 0$ there exists a $K\in \{0,\ldots,N\}$ such that $y\in [c_{k+1}, c_k)$. Then for $i>k+1$ we have $y\geq c_{k+1}\geq c_{i-1}>c_i$ and
\begin{eqnarray*}
\max_{x_{i-1}\leq x< x_i} (U(x)-xy)&=&(c_i-y)x_{i-1}+d_i\\
&=& (c_{i-1}-y)x_{i-1} + d_{i-1}\\
&\leq & (c_{i-1}-y)x_{i-2} + d_{i-1}.
\end{eqnarray*}
The second equality is due to the continuity of $U$ at point $x_{i-1}$
and the third inequality comes from    $c_{i-1}-y\leq 0$ and $x_{i-1}>x_{i-2}$.
We can continue until $i-1=k+1$ to get
$$ \sup_{i>k+1}  \max_{x_{i-1}\leq x< x_i} (U(x)-xy) \leq (c_{k+1}-y)x_{k} + d_{k+1}
=\max_{x_{k}\leq x< x_{k+1}} (U(x)-xy).$$
Similarly, for $i<k$, we have $y<c_k\leq c_{i+1}<c_i$ and
\begin{eqnarray*}
\max_{x_{i-1}\leq x< x_i} (U(x)-xy)&=&(c_i-y)x_{i}+d_i\\
&=& (c_{i+1}-y)x_{i} + d_{i+1}\\
&\leq & (c_{i+1}-y)x_{i+1} + d_{i+1}.
\end{eqnarray*}
The second equality is due to the continuity of $U$ at $x_i$ and the third inequality is from $y\leq c_{i+1}$ and $x_i<x_{i+1}$. We can continue until $i+1=k$ to get
$$ \sup_{i<k}  \max_{x_{i-1}\leq x< x_i} (U(x)-xy) \leq (c_{k}-y)x_{k} + d_{k}
=\max_{x_{k-1}\leq x< x_{k}} (U(x)-xy).$$
From continuity of $U$ at $x_k$ we know that
$\max_{x_{k}\leq x< x_{k+1}} (U(x)-xy)=\max_{x_{k-1}\leq x< x_{k}} (U(x)-xy)$ which is the maximum of $U(x)-xy$ over all intervals if $y\in [c_{k+1},c_k)$, which leads to (\ref{dual-piecewise}).
\qed

We can now characterize the dual value function.
\begin{theorem}
Assume that  $U$ is a piecewise linear utility function defined in (\ref{piecewise}). The dual value function $v(t,y)$ is given by
$$ v(t,y)=\sum_{i=0}^N  A_i(t,y),$$
where, for $i=0,\ldots,N$,
\begin{eqnarray}
A_i(t,y)&=&-x_iy(\Phi(\bar c_{i+1}(y)+\alpha(t))-\Phi(\bar c_i(y) +\alpha(t)))\nonumber\\
&& {}+(c_{i+1}x_i+d_{i+1})(\Phi(\bar c_{i+1}(y))-\Phi(\bar c_i(y)))\label{Ai}\\
\alpha(t)&=&\left(\int_t^T |\hat \theta(u)|^2du\right)^{1/2}\label{alpha}\\
\bar c_i(y)&=&-{1\over \alpha(t)}\ln c_i + {1\over \alpha(t)}\ln y -{1\over 2}\alpha(t)\label{barci}\\
\Phi(x)&=&\int_{-\infty}^x \phi(u)du = \int_{-\infty}^x {1\over \sqrt{2\pi}} e^{-{u^2\over 2}}du.\nonumber
\end{eqnarray}
\end{theorem}

\noindent{\it Proof}.
Consider the linear SDE
$$
d\hat{Y}_s= -\hat{Y}_s\hat{\theta}(s)^T d W_s, \quad s\geq t
$$
with the initial value $\hat{Y}_t=y$. Denote by
$\hat Y^{t,y}_s$ the unique strong solution  and we have
$$ \hat Y^{t,y}_s = y\exp\left(-\int_t^s \hat \theta(u)^T dW_u -{1\over 2}\int_t^s
|\hat \theta(u)|^2du\right).$$
Therefore,
$\hat Y^{t,y}_T=\bar y\exp(-\alpha(t) Z)$,
where
$\alpha(t)$ is given by (\ref{alpha}), $\bar y=y\exp(-\alpha(t)^2/2)$ and $Z$  a standard normal variable. We can find that
$$ v(t,y)=E[\tilde U(\hat{Y}^{t,y}_T)]
=\sum_{i=0}^N \int_{-\infty}^\infty
(-x_i \bar y e^{-\alpha(t) z}+c_{i+1}x_i +d_{i+1}) 1_{\{c_{i+1}\leq \bar y e^{-\alpha(t) z}<c_i\}} \phi(z) dz.$$
Since $c_{i+1}\leq \bar y e^{-\alpha(t) z}<c_i$ is equivalent to
$\bar c_i(y)< z\leq \bar c_{i+1}(y)$, where $\bar c_i(y)$ is defined by (\ref{barci})
and $-\infty=\bar c_0(y)<\bar c_1(y)<\cdots<\bar c_N(y) < \bar c_{N+1}(y)=\infty$. A simple calculation shows that
$v(t,y)=\sum_{i=0}^N  A_i(t,y)$,
where
$A_i(t,y)$ is given by (\ref{Ai})
for $i=0,\ldots,N$.
\qed

We can construct a solution $u(t,x)$ to the HJB equation by first
finding a solution $y(t, x)$ to the equation $v_y(t,y)+x=0$ and then setting $u(t,x)=v(t,y(t,x))+xy(t,x)$.

\begin{proposition}
Consider the wealth maximization problem (\ref{primal}) with $U(x)=x\wedge H$.
Then optimal value function is given by
$$ u(t,x)=v(t, y(t,x))+xy(t,x)
= H\Phi\left(\Phi^{-1}({x\over H}) + \alpha(t)\right)$$
with the optimal feedback control
$$ \pi(t,x)=(\sigma(t)^T)^{-1}\hat\theta(t){H\over x\alpha(t)}\phi\left(\Phi^{-1}({x\over H})\right)$$
and the optimal wealth process
$$X_t=H\Phi(Z_t),$$
where $\alpha(t)$ is given by (\ref{alpha}),
\begin{equation} Z_t=\beta(t)\left( Z_0 + \int_0^t {|\hat \theta(s)|^2\over \alpha(s)\beta(s)}
ds + \int_0^t {1\over \alpha(s)\beta(s)}  \hat \theta(s)^T dW_s\right) \label{Zt}
\end{equation}
and
$\beta(t)=\exp\left(\int_0^t {|\hat \theta(u)|^2\over 2\alpha(u)^2} du\right)$.
\end{proposition}

\noindent{\it Proof}.
Since $U(x)=x\wedge H$ is a piecewise linear function (\ref{piecewise}) with
 $N=1, c_1=1, d_1=0, c_2=0, d_2=H, x_0=0, x_1=H$. We can compute
$\bar c_0(y)=-\infty$ and $\bar c_2(y)=\infty$, which gives
$\bar c_1(y)={1\over \alpha(t)}\ln y-{1\over 2}\alpha(t)$, $A_0(t,y)=0$ and
$ A_1(t,y)=-Hy(1-\Phi(\bar c_1(y)+\alpha(t))) + H(1-\Phi(\bar c_1(y)))$.
Therefore the dual value function is given by
$$ v(t,y)= -Hy\Phi\left(-{1\over \alpha(t)}\ln y-{1\over 2}\alpha(t)\right) +
H\Phi\left(-{1\over \alpha(t)}\ln y+{1\over 2}\alpha(t) \right)$$
and
$$ v_y(t,y)=-H\Phi\left(-{1\over \alpha(t)}\ln y-{1\over 2}\alpha(t)\right).$$
The solution to equation $v_y(t,y)+x=0$ is given by
$$ y(t,x)=\exp\left(-\alpha(t) \Phi^{-1}({x\over H})-{1\over 2}\alpha(t)^2\right).$$
The value function for the primal problem is given by
$$ u(t,x)=v(t, y(t,x))+xy(t,x)
= H\Phi\left(\Phi^{-1}({x\over H}) + \alpha(t)\right).$$
The optimal feedback control is given by
$$ \pi(t,x)=-(\sigma(t)^T)^{-1}\hat\theta(t){u_x(t,x)\over xu_{xx}(t,x)}
=(\sigma(t)^T)^{-1}\hat\theta(t){H\over x\alpha(t)}\phi\left(\Phi^{-1}({x\over H})\right).$$
Substituting the optimal feedback control $\pi(t,x)$ into equation (\ref{wealth}) we get the optimal wealth process $X_t$ satisfies a nonlinear SDE
$$ dX_t= \hat \theta(t)^T(\sigma(t))^{-1} b(t) {H\over \alpha(t)} \phi(\Phi^{-1}({X_t\over H}))dt
+ \hat \theta(t)^T {H\over \alpha(t)} \phi(\Phi^{-1}({X_t\over H})) dW_t.$$
Define $Z_t=f(X_t)$, where $f(x)=\Phi^{-1}({x\over H})$. Ito's lemma implies
that
$$ dZ_t= \left({1\over \alpha(t)} |\hat \theta(t)|^2
+{1\over 2\alpha(t)^2} |\hat \theta(t)|^2 Z_t\right)dt
+\hat \theta(t)^T{1\over \alpha(t)} dW_t.$$
The solution $Z_t$ is given by (\ref{Zt}), which results in
the optimal wealth process $X_t=H\Phi(Z_t)$.
Since $X_t$ is bounded above by the threshold level $H$ the expected utility
of terminal wealth is simply $E[X_T]$.
\qed

\begin{remark}{\rm
Note that if $H$ were set to be infinite then  the problem would be  either trivial with an obvious optimal control  if $K$ is bounded or would be ill-defined with no optimal controls if $K$ is unbounded (the expected portfolio wealth could be made arbitrary large).

A more general utility function than $U(x)=x\wedge H$ is
$$ U_p(x)=\left\{\begin{array}{ll}
x, & 0\leq x< H\\
H(x/H)^p, & x\geq H,
\end{array}\right.
$$
where $H>0$ and $0<p<1$. $U_p$ is a continuous increasing concave function satisfying $U_p(0)=0, U'_p(0)=1$ and $U_p(\infty)=\infty, U'_p(\infty)=0$, $U_p$ is not differentiable at $x=H$ and is not strictly concave on the interval $[0,H]$, and $\lim_{p\downarrow 0} U_p(x)=x\wedge H$.  One may interpret that  $U_p$ is a utility function for an investor who wants to maximize the absolute portfolio wealth up to a threshold $H$ and then a scaled power utility function when the portfolio wealth is more than $H$.

The dual function is defined by $\tilde U_p(y)=\max_{x\geq 0}(U_p(x)-xy)$. By discussing the cases of  $0\leq x\leq H$ and $x\geq H$ we can find the maximum point  $x^*$ and get
$$ \tilde U_p(y)=\left\{\begin{array}{ll}
H{1-p\over p} p^{1\over 1-p} y^{p\over p-1}, &
0<y\leq p\mbox{ with } x^*=H(y/p)^{1\over p-1}\\
H(1-y), & p< y \leq 1\mbox{ with } x^*=H \\
0, & y\geq 1\mbox{ with } x^*=0.
\end{array}\right.
$$
Some long but straightforward calculation shows that the dual value function
is given by
$$
 v_p(t,y)=
H\bigg(  {p_1\over p} p^{1\over p_1} y^{-{p\over p_1}}e^{\alpha(t)^2p\over 2 p_1^2}
\Phi(-c_2+{\alpha(t) p\over p_1})
+\Phi(c_2)-\Phi(c_1)-y\Phi(c_2+\alpha(t))+y\Phi(c_1+\alpha(t)) \bigg),
$$
where $p_1=1-p$, $c_1={1\over \alpha(t)}\ln y - {1\over 2}\alpha(t)$ and
$c_2=c_1 - {1\over \alpha(t)}\ln p$, and its partial derivative  with respect to $y$ is given by
$$
{\partial \over \partial y} v_p(t,y)
=H\left( \Phi(c_1+\alpha(t)) - \Phi(c_2+\alpha(t))
-(y/p)^{1\over p-1}e^{\alpha(t)^2 p\over 2(p-1)^2}\Phi(-c_2+{\alpha(t) p\over 1-p})\right).$$
Finally, one can construct the optimal value function $u(t,x)$ as
$$ u_p(t,x)=v_p(t,y(t,x)) + x y(t,x),$$
where $y(t,x)$ is the unique solution to the equation
${\partial \over \partial y} v_p(t,y)+x=0$.
}\end{remark}

To introduce the concept of risk we may take a trading strategy of investing only in the savings account ($\pi_t\equiv 0$) then
 the terminal wealth is $X_T=x$, the initial endowment (the riskless interest
rate is assumed to be zero). One  has a
certain portfolio wealth $c=U(X_T)=x$ (assuming $x<H$) at time $T$.
To measure the risk associated with a trading strategy $\pi$
 we define a  random variable  $Z$ by
$$Z = c- U(X_T).$$
 $Z$ represents the portfolio wealth loss (or gain) if $Z$ is positive
 (or negative).
Two common risk measures are VaR and CVaR, defined by
\begin{eqnarray*}
 {\rm VaR}_\beta&=&\min\{z: P(Z\leq z)\geq \beta\} \\
{\rm CVaR}_\beta&=&E[Z| Z\geq  {\rm VaR}_\beta],
\end{eqnarray*}
where $\beta\in (0,1)$ (close to 1) is a given constant.
Rockafellar and Uryasev \cite[Theorem 10]{ru02} establish a fundamental representation formula to compute ${\rm VaR}_\beta$
and ${\rm CVaR}_\beta$  by solving  a convex minimization
problem in which the minimum value is ${\rm CVaR}_\beta$
and the left end point
of the minimum solution set gives ${\rm VaR}_\beta$. Specifically,
\begin{equation} \label{CVaR-RU}
{\rm CVaR}_\beta = \min_{y} [ y+ \delta E(Z-y)^+]\label{CVaR}
\end{equation}
where $\delta=(1-\beta)^{-1}$.
If $y^*$ is the left endpoint of the minimum solution set, then
${\rm VaR}_\beta=y^*$.

We may define a new objective function which reconciles  two conflicting objectives
of portfolio wealth  maximization and CVaR minimization. Consider
the following optimization problem
\begin{equation}
\sup_{\pi} (E[U(X_T)] - \lambda {\rm CVaR}_\beta)
\;\mbox{ subject to (\ref{wealth}),} \label{obj1}
\end{equation}
where $\lambda$ is a nonnegative parameter.
 $\lambda=0$ corresponds to the portfolio wealth maximization while
 $\lambda\to\infty$ to the CVaR minimization.
By letting $\lambda$ change in $[0,\infty)$ we can derive the
utility-CVaR efficient frontier in the same spirit as
Markovitz's mean-variance efficient frontier.

Substituting (\ref{CVaR}) into (\ref{obj1}) and exchanging the order of maximization
we may determine the efficient frontier
of  wealth and CVaR by a two-stage optimization problem:
first solving
a parametric utility maximization problem
\begin{equation}
u^y(x) =  \sup_{\pi} E [U^y (X_T)]
\;\mbox{ subject to (\ref{wealth})}, \label{value_y}
\end{equation}
where
\begin{equation}
U^y(x)= U(x) - \lambda \delta(c-U(x) -y)^+, \label{ufunc}
\end{equation}
which is a continuous piecewise linear function and then solving a scalar maximization  problem
\begin{equation}
u(x)=\sup_{y} (u^y(x) - \lambda  y). \label{scalaropt}
\end{equation}
Since $U(x,y)$ is a jointly concave function the optimal value $u^y(x)$ of the first stage optimization problem is a continuous concave function of $y$ and the second stage optimization problem can be easily solved. One only needs to focus on how to solve the first stage optimization
problem.

%provided that the joint optimization problem
%$$
%\sup_{\pi, y} E [U^y(X(T))]
%\;\mbox{ subject to (\ref{wealth})}
%$$
%has a finite optimal value.

\begin{example} \label{example-wealth1}
{\rm
We present a case study to illustrate the wealth maximization, the  CVaR, and the  efficient frontier. Assume that the utility function is $U(x)=x\wedge H$ and the wealth process $X$ follows the process (\ref{wealth}) in which all coefficients are constant and $n=1$.
We know from the above discussion that
$\hat\theta(t)=\theta:=b/\sigma$, $\alpha(t)=\theta \sqrt{T-t}$, and
$\beta(t)=\sqrt{T/(T-t)}$. We can easily find that
$$ Z_t={1\over \sqrt{T-t}} (Z_0 \sqrt{T}+\theta t + W_t)$$
and $Z_0=\Phi^{-1}(x/H)$.
Since $Z_T$ is either $+\infty$ or $-\infty$, depending on the sign of $Z_0\sqrt{T}+\theta
 T +W_T$, we conclude that the optimal terminal wealth $X_T=H$ with probability
 $\Phi(Z_0+\theta \sqrt{T})$ and 0 with probability $\Phi(-Z_0-\theta \sqrt{T})$
 and the expected value of $X_T$ is $H\Phi(Z_0+\theta \sqrt{T})$ which is
the same as  the optimal value  $u(0,x)$ as expected.

It is clear that the Sharp ratio $\theta$ and the investment horizon $T$
have positive impact to both the optimal value and the survival probability.
It is less clear what the level $H$ one should choose. Fox a fixed initial
wealth $x$ the optimal value at time 0 is a function of $H$,
denote by $g(H)$, i.e., $g(H)=H\Phi\left(\Phi^{-1}({x\over H}) + \theta \sqrt{T}\right)$. Since
$$ g'(H)=\Phi\left(\Phi^{-1}({x\over H}) + \theta \sqrt{T}\right)-{x\over H}\exp\left(
-\theta \sqrt{T} \Phi^{-1}({x\over H}) -{1\over 2}\theta^2 T\right)$$
and $g''(H)<0$ and $g'(x)=1, g'(\infty)=0$, which implies that $g$ in an increasing function of $H$. One should set $H$ as large as possible if the objective is to maximize the expected terminal wealth up to a level $H$. On the other hand, the probability of $X_T=0$ increases as $H$ increases,
in other words, the higher the level $H$, the riskier the
portfolio wealth $X_T$,
and in the extreme case when $H=\infty$ the optimal terminal wealth $X_T=0$ almost surely. We conclude that there is no optimal level $H$ for the problem, which
depends on investors' risk preferences.

The CVaR associated with the optimal strategy $\pi^*$
can be computed from the representation (\ref{CVaR-RU}).
Since the optimal terminal wealth $X_T$ is a Bernoulli
variable taking values 0 and $H$, we have
$$
E[(Z-y)^+]=
\left\{\begin{array}{ll}
x-y-HP(X_T=H), & \mbox{if }y<x-H\\
(x-y)P(X_T=0), & \mbox{if }x-H<y<x\\
0, & \mbox{if }y>x.
\end{array}\right.
$$
Solving a simple piecewise linear optimization problem (\ref{CVaR-RU})
we get
$$ {\rm CVaR}_\beta =
\left\{\begin{array}{ll}
x, & \mbox{if }\beta\geq P(X_T=H)\\
x-H(1-\delta P(X_T=0)), & \mbox{if }\beta < P(X_T=H).
\end{array}\right.
$$
For a sufficiently high level of confidence $\beta$ the CVaR is $x$, which
corresponds to the case when $X_T=0$. We also see that when $H\to\infty$
the probability $P(X_T=H)\to 0$ and ${\rm CVaR}_\beta\to x$ for  any
$\beta$, which means one will almost surely lose all initial investment.
This example
illustrates the drawback of simply maximizing the portfolio wealth
without considering the potential loss due to the risky optimal trading strategy.

We now discuss the efficient frontier problem of wealth and CVaR.
The utility function $U^y(x)$ in the first stage problem
is a piecewise linear function for fixed $y$. There are three cases to consider:
\begin{enumerate}
\item If $y<c-H$ then
$ U^y(x)=(1+\lambda\delta)(x\wedge H) - \lambda\delta (c-y)$.
\item If  $c-H\leq y\leq c$ then
$$  U^y(x)=\left\{\begin{array}{ll}
(1+\lambda \delta) x - \lambda \delta (c-y), & 0\geq x<c-y\\
x,& c-y\geq x<H\\
H,&x\geq H
\end{array}\right.
$$
\item If $y> c$ then
$U^y(x)=x\wedge H$.
\end{enumerate}

Cases 1 and 3 are easy and have been discussed above. For case 1 the optimal value is
$$ u^y(x)=(1+\lambda\delta)H\Phi(\Phi^{-1}({x\over H}) + \alpha)
- \lambda\delta (c-y).
$$
The maximum value of $u^y(x)-\lambda y$ over $y\leq c-H$ is reached at point $y=c-H$ and the maximum value is $(1+\lambda\delta)H\Phi(\Phi^{-1}({x\over H}) + \alpha)
- \lambda\delta H$.
For case 3 the optimal value is
$$ u^y(x)=H\Phi(\Phi^{-1}({x\over H}) + \alpha).
$$
The maximum value of $u^y(x)-\lambda y$ over $y\geq c$ is reached at point $y=c$.

Since $u^y(x)$ is a continuous function of $y$ we know that the maximum of $u^y(x)-\lambda y$ over $y\in R$ is achieved in the interval
$[c-H,c]$, which corresponds to case 2.

Consider a piecewise linear utility function
$$ U(x)=\left\{\begin{array}{ll}
(1+k)x-kh, & 0\leq x<h \\
x, & h\leq x<H\\
H, & H\leq x.
\end{array}\right.
$$
The case 2 corresponds to $h=c-y$ and $k=\lambda \delta$.

This is a special case of a piecewise linear function defined in (\ref{piecewise}). The data are $N=2, c_1=1+k, d_1=-kh, c_2=1, d_2=0, c_3=0, d_3=H,
x_0=0, x_1=h, x_2=H$, and $\bar c_0(y)=-\infty, \bar c_3(y)=\infty$,
where $h=c-y$ and $k=\lambda \delta$.
We have
$\bar c_1(y)=-{1\over \alpha} \ln(1+k) +{1\over \alpha}\ln y - {1\over 2}\alpha$,
$\bar c_2(y)={1\over \alpha} \ln y - {1\over 2}\alpha$, and
$$ v(t,y)=A_0(t,y)+A_1(t,y)+A_2(t,y),$$
where
\begin{eqnarray*}
 A_0(t,y)&=&-kh \Phi(\bar c_1(y)),\\
A_1(t,y)&=&-hy(\Phi(\bar c_2(y)+\alpha)-\Phi(\bar c_1(y)+\alpha)) + h(\Phi(\bar c_2(y))-\Phi(\bar c_1(y))), \\
A_2(t,y)&=&-Hy(1-\Phi(\bar c_2(y)+\alpha)) + H(1-\Phi(\bar c_2(y))).
\end{eqnarray*}
A simple calculus shows that
$$ v_y(t,y)=-H +h\Phi(\bar c_1(y)+\alpha) + (H-h)\Phi(\bar c_2(y)+\alpha).$$
There is no closed form solution $y(t,x)$ to the equation $g(y):=v_y(t,y)+x=0$.
Since $g(0+)=-H+x<0$ and $g(\infty)=x>0$ and $g$ is strictly increasing and there exsits a unique root to the equation $g(y)=0$. In general, one has to use some numerical method to find the unique root.
}\end{example}

%===================================
% turnpike
%===================================

\section{Turnpike Property}

In this section we discuss turnpike property when $T\rightarrow \infty$.
This property is studied in \cite{HZ99}. Here we give a
new and direct proof with the PDE approach and the duality method.
We also improve the results of \cite{HZ99}.

In this section we assume that the market is made up of one riskless asset with interest rate $r$ and one risky asset with price $S$ satisfying the SDE $dS=\mu Sdt+ \sigma S dW$, where $r,\mu,\sigma$ are constant satisfying  $\mu>r>0$.
Assume that the utility function $U$ satisfies Assumption \ref{utility} and $U(\infty)=\infty$.
 Let
$u(t,x;T)=\sup E[U(X_T)|X_t=x]$ be the value function and the solution to the HJB equation and let $\bar u(\tau,x)=u(t,x;T)$  with $\tau=T-t$,  the time to maturity. We may continue to write $u$ instead of $\bar u$ but with $\tau$ as time-to-maturity variable.
Theorem \ref{maintheorem} (or  Theorem 3.8 in \cite{bmz11}) says that $u$ is a classical solution to the HJB equation satisfying
\begin{equation}
-u_\tau-\frac{1}{2}\theta^2\frac{u_x^2}{u_{xx}}+rxu_x=0,\quad (\tau,x)\in R_+\times R_+
\end{equation}
with the initial condition $u(0,x)=U(x)$ for $x\in R_+$, where $\theta=\frac{\mu-r}{\sigma}$.
We see that $u\in C^{1,\infty}(R_+\times R_+)$ and $u_{xx}<0$ for $\tau>0$.
The optimal policy (the amount invested in the risky asset) is given by
\[
A(\tau,x)=- {\theta\over \sigma} \frac{u_x(\tau,x)}{u_{xx}(\tau,x)},\quad \tau>0
\]
is the solution of the following equation
\[
A_\tau-\frac{1}{2}\theta^2 A^2 A_{xx}-rxA_x+rA=0,(\tau,x)\in R_+\times R_+.
\]

Let $\tilde U(y)$ be the dual of $U(x)$. Then the dual $v(\tau,x)$ of $u(\tau,x)$ satisfies
\begin{equation}
v_\tau-\frac{1}{2}\theta^2 y^2 v_{yy}+ryv_y=0,(\tau,y)\in R_+\times R_+
\end{equation}
with the initial condition $v(0,y)=\tilde U(y)$ for $y\in R_+$ and $v\in C^{1,\infty}$.
It is easy to verify that $w=v_y$ and $w=yv_{yy}$
are the solutions to the equation
\begin{equation}
Lw:=w_\tau-\frac{1}{2}\theta^2 y^2 w_{yy}+(r-\theta^2)yw_y+rw=0,(\tau,y)\in R_+\times R_+.
\end{equation}

\begin{lemma} \label{Asy}
Assume that $w\in C^{1,2}$ is a solution to the linear parabolic equation
\[
w_\tau-a^2 w_{xx}=0,(\tau,x)\in R\times R_+
\]
with the initial condition $w(0,x)=\phi(x)$ for $x\in R$.
Assume that $\phi(x)\in C(R)$ and
\begin{equation}\label{C1}
\lim_{x\rightarrow \infty} e^{\alpha x} \phi(x)=0, \lim_{x\rightarrow -\infty}\frac{\phi(x)}{e^{kx}}=1,
\end{equation}
where $\alpha>0$ and $k+\alpha<0$. Then we have
\begin{equation}\label{PA}
\lim_{x\rightarrow \infty}e^{\alpha x}w(\tau,x)=0,\quad
\lim_{x\rightarrow \infty}e^{\alpha x}w_x(\tau,x)=0,\quad
\lim_{x\rightarrow \infty}e^{\alpha x}w_{xx}(\tau,x)=0,
\end{equation}
and
\begin{equation}\label{NA1}
\lim_{x\rightarrow -\infty} {w(\tau,x) \over e^{k^2 a^2 \tau+kx}}=1,\quad
\lim_{x\rightarrow -\infty} {w_x(\tau,x)\over e^{k^2 a^2 \tau+kx}}=k, \quad
 \lim_{x\rightarrow -\infty} {w_{xx}(\tau,x)\over e^{k^2 a^2 \tau+kx}}=k^2.
\end{equation}
The convergence is uniform for $\tau\in[\tau_0,\tau_1]$ for any $0<\tau_0<\tau_1$.
\end{lemma}

\noindent{\it Proof.}\
We first prove (\ref{PA}). From (\ref{C1}), there exists constant $C_0$, such that
\[
e^{\alpha x}\phi(x)\leq C_0(1+e^{(\alpha+k)x}), \forall x.
\]
By Poisson's formula we have
\begin{eqnarray*}
w(\tau,x)=\frac{1}{2\sqrt{\pi }}\int_{-\infty}^\infty
e^{-\frac{\eta^2}{4}}\phi(x+a\sqrt{\tau}\eta)d\eta.
\end{eqnarray*}
Hence,
\begin{eqnarray*}
e^{\alpha x}w(\tau,x)&=&\frac{1}{2 \sqrt{\pi}}\int_{-\infty}^\infty
e^{-\frac{\eta^2}{4}-\alpha a\sqrt{\tau}\eta}[e^{\alpha(x+a\sqrt{\tau}\eta)}\phi(x+a\sqrt{\tau}\eta)]d\eta\\
&\leq& \frac{C_0}{2 \sqrt{\pi }}\int_{-\infty}^{-A}
e^{-\frac{\eta^2}{4}-\alpha a\sqrt{\tau}\eta}(1+e^{(\alpha+k)(x+a\sqrt{\tau}\eta)})d\eta\\
&& {}+\frac{1}{2\sqrt{\pi }}\int_{-A}^\infty
e^{-\frac{\eta^2}{4}-\alpha a\sqrt{\tau}\eta}[e^{\alpha (x+a\sqrt{\tau}\eta)}\phi(x+a\sqrt{\tau}\eta)]d\eta.
\end{eqnarray*}
For any $\epsilon>0$, there exists $A=A(\epsilon)>0$, such that, for $x>0$
\[
\frac{C_0}{2 \sqrt{\pi }}\int_{-\infty}^{-A}
e^{-\frac{\eta^2}{4}-\alpha a\sqrt{\tau}\eta)}(1+e^{(\alpha+k)a\sqrt{\tau}\eta})d\eta <\epsilon.
\]
For $\eta\geq -A$, we see that $x+a\sqrt{\tau}\eta\geq x-a\sqrt{\tau}A$. Hence there exists $X>0$, such that
\[
\frac{1}{2\sqrt{\pi }}\int_{-A}^\infty
e^{-\frac{\eta^2}{4}-\alpha a\sqrt{\tau}\eta)}[e^{\alpha (x+a\sqrt{\tau}\eta)}\phi(x+a\sqrt{\tau}\eta)]d\eta <\epsilon
\]
if $x>X$. This proves the first limit in (\ref{PA}).
The other two limits  in (\ref{PA}) are proved similarly with the help of Poisson's formula and its derivatives with respect to $x$.

We next prove (\ref{NA1}). Since
\[
\frac{w(\tau,x)}{e^{k^2 a^2 \tau+kx}}=\frac{1}{2 \sqrt{\pi}}\int_{-\infty}^\infty
e^{-\frac{(\eta-2ka\sqrt{\tau})^2}{4}}\frac{\phi(x+a\sqrt{\tau}\eta)}{e^{k(x+a\sqrt{\tau}\eta)}}d\eta
\]
we have
\begin{eqnarray*}
|\frac{w(\tau,x)}{e^{k^2 a^2 \tau+kx}}-1|&\leq& \frac{1}{2 \sqrt{\pi}}\int_{-\infty}^\infty
e^{-\frac{(\eta-2qa\sqrt{\tau})^2}{4}}|\frac{\phi(x+a\sqrt{\tau}\eta)}{e^{k(x+a\sqrt{\tau}\eta)}}-1|d\eta\\
&\leq& \frac{1}{2 \sqrt{\pi}}\int_{-\infty}^A
e^{-\frac{(\eta-2qa\sqrt{\tau})^2}{4}}|\frac{\phi(x+a\sqrt{\tau}\eta)}{e^{k(x+a\sqrt{\tau}\eta)}}-1|d\eta\\
&& {}+\frac{C_0}{2\sqrt{\pi}}\int_{A}^\infty
e^{-\frac{(\eta-2qa\sqrt{\tau})^2}{4}}[1+\frac{1}{C_0}+e^{-(\alpha+k)(x+a\sqrt{\tau}\eta)}]d\eta.
\end{eqnarray*}
For any $\epsilon>0$, there exists $A=A(\epsilon)>0$, such that, for $x<0$
\[
\frac{C_0}{2\sqrt{\pi}}\int_{A}^\infty
e^{-\frac{(\eta-2qa\sqrt{\tau})^2}{4}}[1+\frac{1}{C_0}+e^{-(\alpha+k)(x+a\sqrt{\tau}\eta)}]d\eta <\epsilon.
\]
For $\eta\leq A$, we see that $x+a\sqrt{\tau}\eta\leq x+a\sqrt{\tau}A$. Hence there exists $X>0$, such that
\[
\frac{1}{2 \sqrt{\pi}}\int_{-\infty}^A
e^{-\frac{(\eta-2qa\sqrt{\tau})^2}{4}}|\frac{\phi(x+a\sqrt{\tau}\eta)}{e^{k(x+a\sqrt{\tau}\eta)}}-1|d\eta <\epsilon
\]
if $x<-X$. This proves the first limit in (\ref{NA1}). The other two limits in (\ref{NA1}) are proved similarly.
\qed

\begin{corollary}\label{cor1}
Assume that  $\tilde U(y)\in C(R_+)$ and
\begin{equation}\label{C2}
\lim_{y\rightarrow \infty}\tilde U(y)=0,\quad \lim_{y\rightarrow 0}\frac{\tilde U(y)}{y^q}=1,
\end{equation}
where $q<0$. Then we have
\begin{equation}\label{PAA}
\lim_{y\rightarrow \infty}v(\tau,y)=0,\lim_{y\rightarrow \infty}yv_y(\tau,y)=0,\lim_{y\rightarrow \infty}y^2v_{yy}(\tau,y)=0
\end{equation}
and
\begin{equation}\label{NAA2}
\lim_{y\rightarrow 0}\frac{v(\tau,y)}{e^{\lambda \tau}y^q}=1,\quad
\lim_{y\rightarrow 0}\frac{v_y(\tau,y)}{e^{\lambda \tau}y^{q-1}}=q,\quad
\lim_{y\rightarrow 0}\frac{v_{yy}(\tau,y)}{e^{\lambda \tau}y^{q-2}}=q(q-1),
\end{equation}
where $\lambda=\frac{1}{2}\theta^2q(q-1)-rq$. The convergence is uniform for $\tau\in[\tau_0,\tau_1]$ for any $0<\tau_0<\tau_1$.
\end{corollary}

\noindent{\it Proof.}\ We have
\begin{equation}
v_\tau-\frac{1}{2}\theta^2 y^2 v_{yy}+ryv_y=0,(\tau,y)\in R_+\times R_+
\end{equation}
with the initial condition $v(0,y)=\tilde U(y)$ for $y\in R_+$.
Let
\[
w(\tau,x)=e^{-(\alpha x+\beta \tau)}v(e^x,\tau)
\]
with $\alpha=\frac{1}{2}+\frac{r}{\theta^2}$ and
$\beta=-\frac{(\theta^2+2r)^2}{8 \theta^2}$.
Then
\[
w_\tau-\frac{1}{2}\theta^2 w_{xx}=0,\quad (\tau,x)\in R\times R_+
\]
with the initial condition $w(0,x)=\phi(x)=e^{-\alpha x}\tilde U(e^x)$ for $x\in R$.
Applying lemma \ref{Asy}, we complete the proof.
\qed

\begin{lemma}\label{UtoV}
Let $U$ be a continuous increasing concave function on $R_+$ satisfying $U(0)=0$ and
\begin{equation}\label{C3}
\lim_{x\rightarrow \infty}\frac{U(x)}{x^p}=\frac{1}{p^p(1-p)^{1-p}},
\end{equation}
where $0<p<1$ is a constant. Then condition (\ref{C2}) holds with $q={p\over p-1}$.
\end{lemma}

\noindent{\it Proof.}\
Since $U$ is a concave function on $R_+$, the superdifferential of $U$ at $x\in R_+$ is a convex compact set, defined by
$$ \partial U(x)=\{\xi: U(y)\leq U(x)+\xi(y-x),\; y\in R_+\}.$$
The increasing property of $U$ and condition (\ref{C3}) imply that $\partial U(x)$ is a subset of $R_+$ and $0\not\in \partial U(x)$ for all $x\in R_+$. (If $0\in \partial U(\bar x)$ for some $\bar x\in R_+$, then  $U(x)\equiv U(\bar x)$ for all $x\geq \bar x$, which contradicts  (\ref{C3}).)
Denote by $a=\frac{1}{p^p(1-p)^{1-p}}$.
By (\ref{C3}), for any $0<\epsilon<a$,
there exists $X_\epsilon>\frac{1}{\epsilon}$, such that
\[
(a-\epsilon)x^p\leq U(x) \leq (a+\epsilon)x^p
\]
for $x\geq X_\epsilon$. The superdifferential of $U$ at $X_\epsilon$ is given by
$\partial U(X_\epsilon)=[a_\epsilon, b_\epsilon]$ for some $0<a_\epsilon\leq b_\epsilon<\infty$. For $y<a_\epsilon$, we have, for $x\leq X_\epsilon$, that
$$ U(x)-xy\leq U(X_\epsilon)+a_\epsilon(x-X_\epsilon)-xy\leq U(X_\epsilon) - X_\epsilon y,$$
which implies that
\begin{eqnarray*}
\tilde U(y)&=&\max_{x\geq X_\epsilon}\{U(x)-xy\}\\
&\leq& \max_{x\geq X_\epsilon}\{(a+\epsilon)x^p-xy\}\\
&\leq& \max_{x\geq 0}\{(a+\epsilon)x^p-xy\}\\
&=&-\frac{1}{q}[p(a+\epsilon)]^{-\frac{1}{p-1}}y^q.
\end{eqnarray*}
On the other hand, if $y< p(a-\epsilon)X_\epsilon^{p-1}$ then function $(a-\epsilon)x^p-xy$ achieves its maximum
on $[X_\epsilon,\infty)$ at $x=(\frac{y}{p(a-\epsilon)})^{\frac{1}{p-1}}$, therefore
\begin{eqnarray*}
\tilde U(y)&=&\max_{x\geq X_\epsilon}\{U(x)-xy\}\\
&\geq& \max_{x\geq X_\epsilon}\{(a-\epsilon)x^p-xy\}\\
&=&-\frac{1}{q}[p(a-\epsilon)]^{-\frac{1}{p-1}}y^q.
\end{eqnarray*}
Dividing $y^q$ on the above inequalities and letting $y\to 0$ and then $\epsilon\rightarrow 0$ we get (\ref{C2}). ($\tilde U(\infty)=0$ is trivial as $U(0)=0$.)
\qed

\begin{theorem} \label{Main}
Assume that $U$ is a continuous increasing concave function on $R_+$ satisfying $U(0)>-\infty$ and
$$  \lim_{x\rightarrow \infty}\frac{U(x)}{x^p}= k,$$
where $0<p<1$ and $k>0$ are constant. Then for any $x\in R_+$ we have
\[
\lim_{\tau\rightarrow \infty}A(\tau,x)=\frac{\theta}{\sigma(1-p)}x.
\]
\end{theorem}

\noindent{\it Proof.}\
Since we are studying the behaviour of the optimal controls which
are invariant for utility functions $U$ and $cU+d$ with $c$ a positive constant and $d$ a constant, we may assume without loss of generality that $U$ satisfies $U(0)=0$ and  (\ref{C3}), which then implies (\ref{C2}) (Lemma \ref{UtoV}) and the results of Corollary \ref{cor1}.
From (\ref{PAA}) and (\ref{NAA2}) we get
\[
\lim_{y\rightarrow 0}\frac{v_{y}(1,y)-qy^{q-1}e^{\lambda}}{y^{q-1}}=0, \quad
\lim_{y\rightarrow \infty}(v_{y}(1,y)-qy^{q-1}e^{\lambda})=0.
\]
For any fix $\epsilon>0$ there is $\delta=\delta(\epsilon)>0$ such that
\[
|v_{y}(1,y)-qy^{q-1}e^{\lambda}|\leq \epsilon y^{q-1}+\delta, y\in R_+.
\]
Let, for $(\tau,y)\in [1,\tau_1] \times R_+$,
\[
w(\tau,y)=\pm (v_{y}(\tau,y)-qy^{q-1}e^{\lambda \tau})+\epsilon (y^{q-1}e^{\lambda (\tau-1)}+1)+\delta e^{-r(\tau-1)}.
\]
Then $w$ satisfies the equation $Lw=r\epsilon$ for $(\tau,y)\in R_+\times R_+$ and
\[
w(1,y)>0,\quad
\liminf_{y\rightarrow 0}w(\tau,y)>0,\quad \liminf_{y\rightarrow \infty}w(\tau,y)>0.
\]
By the maximum principle we conclude that $w(\tau,y)>0$, which gives
\[
|v_{y}(\tau,y)-qy^{q-1}e^{\lambda \tau}|\leq \epsilon (y^{q-1}e^{\lambda (\tau-1)}+1)+\delta e^{-r(\tau-1)}
\]
for all $(\tau,y)\in  [1,\tau_1] \times R_+$. Hence
\[
|v_{y}(\tau,u_x(\tau,x),\tau)-q(u_x(\tau,x))^{q-1}e^{\lambda \tau}|\leq \epsilon ((u_x(\tau,x))^{q-1}e^{\lambda (\tau-1)}+1)+\delta e^{-r(\tau-1)}.
\]
Since $v_{y}(u_x(\tau,x))=-x$ we obtain
\[
|q(u_x(\tau,x))^{q-1}e^{\lambda \tau}|\leq |x|+\epsilon ((u_x(\tau,x))^{q-1}e^{\lambda (\tau-1)}+1)+\delta e^{-r(\tau-1)}.
\]
Taking $\epsilon=-\frac{q}{2}e^{\lambda}$, we conclude that
there is constant $C=C(x)$, such that
\begin{equation}\label{Es1}
|(u_x(\tau,x))^{q-1}e^{\lambda \tau}|\leq C(x).
\end{equation}

Similarly, from (\ref{PAA}) and (\ref{NAA2}) we get
\[
\lim_{y\rightarrow 0}\frac{yv_{yy}(1,y)+(1-q)v_y(1,y)}{y^{q-1}}=0,\quad \lim_{y\rightarrow \infty}(yv_{yy}(1,y)+(1-q)v_y(1,y))=0.
\]
For any fix $\epsilon>0$, there is $\delta=\delta(\epsilon)>0$, such that
\[
|yv_{yy}(1,y)+(1-q)v_y(1,y)|\leq \epsilon y^{q-1}+\delta, y\in R_+.
\]
Define
\[
w(\tau,y)=\pm (yv_{yy}(\tau,y)+(1-q)v_y(\tau,y))+\epsilon (y^{q-1}e^{\lambda (\tau-1)}+1)+\delta e^{-r(\tau-1)}
\]
for $(\tau,y)\in  [1,\tau_1] \times R_+$. Then $w$ satisfies the equation
$Lw=r\epsilon$ and
\[
w(1,y)>0,\quad
\liminf_{y\rightarrow 0}w(\tau,y)>0,\quad
\liminf_{y\rightarrow \infty}w(\tau,y)>0.
\]
By the maximum principle we conclude that $w(\tau,y)>0$, hence
\[
|yv_{yy}(\tau,y)+(1-q)v_y(\tau,y)|\leq \epsilon (y^{q-1}e^{\lambda (\tau-1)}+1)+\delta e^{-r(\tau-1)}
\]
for all $(\tau,y)\in  [1,\tau_1] \times R_+$.
Now fix $x$, by (\ref{Es1}), we obtain
\begin{eqnarray*}
|A(\tau,x)-\frac{\theta}{\sigma(1-p)}x| &=& {\theta\over\sigma} |u_x(\tau,x)v_{yy}(\tau,u_x(\tau,x))+(1-q)v_y(\tau,u_x(\tau,x))|\\
&\leq& {\theta\over\sigma}\left(\epsilon ((u_x(\tau,x))^{q-1}e^{\lambda (\tau-1)}+1)+\delta e^{-r(\tau-1)}\right)\\
&\leq& {\theta\over\sigma}\left(\epsilon (C(x)+1)+\delta e^{-r(\tau-1)}\right).
\end{eqnarray*}
This yields
\[
\limsup_{\tau\rightarrow \infty}|A(\tau,x)-\frac{\theta}{\sigma(1-p)}x|\leq  {\theta\over\sigma}\epsilon (C(x)+1),
\]
which implies that
\[
\lim_{\tau\rightarrow \infty}A(\tau,x)=\frac{\theta}{\sigma(1-p)}x.
\]
\qed

\section{Conclusion}
In this paper we extend the results of  \cite{bmz11} to a more general class  of utility functions which may be bounded and not necessarily  strictly concave and show that there is a classical solution to the HJB equation with the dual control method. We then apply the results to two specific problems: One is the efficient frontier of wealth and CVaR and the other the turnpike property. For the first problem we construct explicitly  the optimal control and discuss the choice of the optimal threadshold level  and conclude that the wealth and the risk (CVaR) are positively correlated and the efficient frontier depends on the risk preference of investors. For the second problem we prove the turnpike property of the optimal policy of long-run investors  by applying the smoothness of the primal and dual value functions, their dual relationship, and the maximum principle of linear parabolic PDEs, which generalizes the results and simplifies the proofs  of  \cite{HZ99}.

\end{document}